\documentclass[twocolumn,aps,prb,epsf,graphics,psfig]{revtex4}
\usepackage{graphicx,graphics} 
\usepackage[dvips,bookmarks=false]{hyperref}

\usepackage{graphicx}
\usepackage{dcolumn}

\usepackage{eucal}
\usepackage[dvips]{epsfig}
\begin{document}
\title{Chemical and magnetic impurity effects on electronic properties of semiconductor
quantum wires}
\author{Alireza Saffarzadeh}
\altaffiliation{E-mail: a-saffar@tehran.pnu.ac.ir}
\affiliation{$^1$Department of Physics,
Payame Noor University, Nejatollahi St., 159995-7613 Tehran, Iran \\
$^2$Computational Physical Sciences Laboratory, Department of
Nano-Science, Institute for Research in Fundamental Sciences
(IPM), P.O. Box 19395-5531, Tehran, Iran}
\date{\today}
\begin{abstract}
We present a theoretical study of electronic states in magnetic
and nonmagnetic semiconductor quantum wires. The effects of
chemical and magnetic disorder at paramagnetic temperatures are
investigated in single-site coherent potential approximation. It
is shown that the nonmagnetic impurity shifts the band of carriers
and suppresses the van Hove singularities of the local density of
states (LDOS) depending on the value of impurity concentration.
The magnetic impurity, however, broadens the band which depends on
the strength of exchange coupling, and in the high impurity
concentration, the van Hove singularities in the LDOS can
completely disappear and the curves become smooth.
\end{abstract}
\maketitle

\section{\bf Introduction}
The nature of the dimensionality and of the confinement associated
with a particular nanostructure such as a quantum well, quantum
wire, or quantum dot have a pronounced effect on its physical
properties. Quantum wires and quantum dots are under active
experimental investigation because devices based on them offer
important opportunities as the building blocks for the next
generation of electronic and opto-electronic devices ranging from
ultrafast optical switching to ultradense memories \cite{Uskov}.
Quantum wire structures have density of states features which are
very useful for laser applications with possibility of smaller
current threshold density than in lasers produced from higher
dimensional structures. However, it is clear that disorder affects
these attractive features of quantum wires
\cite{Taylor,Nik0,Harris}. Over the past decade, using the
coherent potential approximation (CPA), the effects of boundary
roughness and the presence of islands on the electronic properties
of nonmagnetic semiconductor (NMS) quantum wires have been studied
\cite{Nik1,Nik2,Hong}. Furthermore, it has been demonstrated by
Ohno \cite{YOhno} that it is indeed possible to inject spin from a
diluted magnetic semiconductor (DMS) to a NMS, which is necessary
in order to carry out qubit (quantum bit) operations required for
quantum computing \cite{DiVince}.

DMS's \cite{Chapman} are semiconductors of the general type
$\mathrm{A}_{1-x}\mathrm{M}_x\mathrm{B}$, where AB is either a
II-VI or a III-V semiconductor and M a magnetic element, most
commonly Mn . Substitution of a small fraction $x$ of the element
A by Mn impurities (and in the case of II-VI semiconductors an
additional charge dopant, such as P on the B site) leads to the
appearance of a semiconductor with ferromagnetic properties
\cite{Furdyna,Ohno}. The magnetic state in these materials has
been attributed to the exchange interaction of the localized Mn
moments with the spin of the charge carriers introduced by the Mn
dopants, or in the case of II-VI semiconductors, by the additional
dopant.

In recent years, considerable works have been devoted to the
understanding of physical properties of DMS quantum wires, both
theoretically \cite{Kim,Kyry,Chang,Xu,Moradian} and experimentally
\cite{Chen1,Chen2,Jeon}. One of the most important physical
quantities in these quantum structures, is the density of states
of charge carriers which depends on the dimensionality of the
structure. The behavior of this quantity is very important in
determining the electrical, thermal, and other properties of the
system. Furthermore, DMS's belong to the class of disordered
systems, hence, any first principle consideration should take
into account the randomness of impurities. Among different kinds
of randomness, two kinds of them can strongly affect the transport
properties of charge carriers in DMS quantum wires, i.e., the
random substitution of the magnetic atoms and the random
direction of the impurity spins.

The aim of this work is to investigate the effects of chemical
(spin-independent) potential and magnetic disorder on the
electronic properties of semiconductor quantum wires. Based on the
single-site CPA for the magnetic and nonmagnetic impurities at
paramagnetic temperatures \cite{Taka99}, we study the local
density of states (LDOS) for charge carriers, in terms of the
impurity concentrations and spin scattering strengths.

The paper is organized as follows. The model, Hamiltonian and
formalism are given in section II. In section III, we present the
results of the numerical calculations for the NMS and DMS quantum
wires. A brief conclusion is given in section IV.

\section{\bf Model and method}
We consider a semiconductor quantum wire described by the
tight-binding model on a square lattice in which one of the
dimensions (the $x$ direction) is much larger than the other (the
$y$ direction), i.e., a long-strip lattice in two dimension. The
sites of the lattice are denoted by ($m$,$n$) where $m$ is an
integer number and $N_y$ is number of atoms in the $y$ direction,
hence $1\leq n\leq N_y$. In fact, we have divided the wire into
$N_y$ atomic lines along the $y$ axis and each line lies along the
$x$ axis. We set the site energies to be infinite along the lines
$n=0$ and $n=N_y+1$; thus, the carriers are confined along the $y$
direction. The one-electron Hamiltonian for this system is given
by
\begin{equation}\label{1}
H=\sum_{i}u_i(\mathrm{M,A})+\sum_{i,j,\sigma}t_{ij}\mid
i,\sigma\rangle\langle j,\sigma\mid \ ,
\end{equation}
where $u_i$ depends on whether $i\equiv(m,n)$ is a magnetic (M) or
nonmagnetic (A) site. For the A-site
\begin{equation}\label{2}
u_i^\mathrm{A}=\varepsilon_\mathrm{A}\sum_{\sigma}\mid
i,\sigma\rangle\langle i,\sigma\mid\  ,
\end{equation}
and for the M-site
\begin{equation}\label{3}
u_i^\mathrm{M}=\sum_{\sigma,\sigma'}\mid
i,\sigma\rangle[\varepsilon_\mathrm{M}\delta_{\sigma\sigma'}-I{\bf
S}_i{\bf\cdot\tau}_{\sigma\sigma'}]\langle i,\sigma'\mid\  ,
\end{equation}
In the above equations, $\mid i,\sigma\rangle$ is an atomic
orbital with spin $\sigma$ (=$\uparrow$ or $\downarrow$) at site
($m$,$n$), $\varepsilon_\mathrm{A}$ and $\varepsilon_\mathrm{M}$
are the on-site energies for A- and M-sites, and the hopping
energy $t_{ij}=t$ if $i$ and $j$ are nearest neighbors and zero
otherwise. The second term on the right hand side of Eq. \ref{3}
is the $\mathbf{k}$-independent exchange interaction in which
${\bf S}_i$ is the local spin operator of the $\mathrm{M}$-atom
and ${\bf\tau}$ is the usual Pauli matrix for carrier's spin. We
regard the spin of magnetic ion as a classical spin, while the
value of exchange interaction strength $IS$ (=$I\times S$) is
finite.

The single-electron Hamiltonian can be written as
\begin{equation}\label{4}
H=\mathcal{H}_{eff}+V\  ,
\end{equation}
where the effective Hamiltonian $\mathcal{H}_{eff}$, which
describes the effective medium, is expressed as
\begin{equation}\label{5}
\mathcal{H}_{eff}=\sum_{i,j,\sigma}[t_{ij}+\delta_{i,j}
\Sigma_{i}(\omega)]\mid i,\sigma\rangle\langle j,\sigma\mid\ .
\end{equation}
Here, $\Sigma_{i}(\omega)$ is the site-dependent self-energy, and
the perturbation term is given as
\begin{eqnarray}\label{6}
V&=&H-\mathcal{H}_{eff} \nonumber \\
&=&\sum_{i}v_i \ ,
\end{eqnarray}
where $v_i=v_i^\mathrm{A}$ for the A-site and $v_i=v_i^\mathrm{M}$
for the M-site are given by
\begin{equation}\label{7}
v_i^\mathrm{A}=\sum_{\sigma}\mid
i,\sigma\rangle[\varepsilon_\mathrm{A}-\Sigma_i]\langle
i,\sigma\mid\ ,
\end{equation}
\begin{equation}\label{8}
v_i^\mathrm{M}=\sum_{\sigma\sigma'}\mid
i,\sigma\rangle[(\varepsilon_\mathrm{M}-\Sigma_i)\delta_{\sigma,\sigma'}-I{\bf
S}_i{\bf\cdot\tau}_{\sigma\sigma'}]\langle i,\sigma'\mid\ ,
\end{equation}

It should be emphasized that, we have assumed a spin-independent
effective medium, since the system is at the paramagnetic
temperatures ($T\gg T_c$). Thus, $\Sigma_i$ does not depend on the
spin of carriers and hence, the electronic states will be
independent of the temperature within the paramagnetic regime.
Here, $T_c$ is defined as the ferromagnetic critical temperature
where the spontaneous magnetization of magnetic impurities
vanishes, and in $\mathrm{Ga}_{1-x}\mathrm{Mn}_x\mathrm{As}$
quantum wires, for example, $T_c$ as high as 350 K has been
reported \cite{Jeon}.

The matrix elements of effective Green's function, $\bar{G}$, can
be determined from the Dyson equation:
\begin{equation}\label{9}
\bar{G}_{i,i'}(\omega)=G^0_{i,i'}(\omega)+
\sum_{j}G^0_{i,j}(\omega)\Sigma_{j}(\omega)\bar{G}_{j,i'}(\omega)\
,
\end{equation}
where ${G}^0_{i,j}$ is the clean system Green's function matrix
element and is given by
\begin{eqnarray}\label{10}
G^0_{i,i'}&=&\langle m,n|G^0(\omega)|m',n'\rangle \nonumber \\
&=&\frac{1}{N_x}\sum_{k_x}\sum_{l=1}^{N_y}\frac{f_{n,n'}(l)}{\omega+i\eta-\epsilon_{l,k_x}}
e^{ik_x(m-m')a}\  ,
\end{eqnarray}
where
\begin{equation}\label{11}
f_{n,n'}(l)=\frac{2}{N_y+1}\sin(\frac{l\pi}{N_y+1}n)\sin(\frac{l\pi}{N_y+1}n')
\  ,
\end{equation}
and
\begin{equation}\label{12}
\epsilon_{l,k_x}=2t\cos(\frac{l\pi}{N_y+1})+2t\cos(k_xa)\  ,
\end{equation}
is the clean system band structure.

Here, $N_x$ and $k_x$ are the number of lattice sites and the wave
vector in the $x$ direction, $a$ is the lattice constant, $l$ is
the mode of the subband, and $\eta$ is a positive infinitesimal.
Since the translational symmetry is absent in the $y$ direction,
the self-energy depends on the atomic line number ($n$), however,
it is independent of the number ($m$) of the atomic site on each
line, i.e. $\Sigma_i(\omega)\equiv\Sigma_n(\omega)$. In this case,
the Dyson equation (\ref{9}) can be rewritten as
\begin{eqnarray}\label{13}
\bar{G}_{n_1,n_2}(m_1,m_2;\omega)=G^0_{n_1,n_2}(m_1,m_2;\omega)~~~~~~~~~~~~~ \nonumber \\
+\sum_{n=1}^{N_y}\sum_{m}G^0_{n_1,n}(m_1,m;\omega)\Sigma_{n}(\omega)
\bar{G}_{n,n_2}(m,m_2;\omega)\ .&&
\end{eqnarray}

The CPA replaces the real system with an effective periodic medium
\cite{Soven,Gonis}. For this purpose, the potential of all sites
is replaced by an energy-dependent coherent potential, except one
site which is denoted by \emph{impurity}. The effective medium is
determined self-consistently in such a way that the Green's
function of the effective medium is equal to the configurationally
averaged Green's function of the real medium. Therefore, the
effective scattering of a carrier at the impurity site is zero, on
average.

In order to derive the CPA equation for the coherent potential, we
write the Green's function of the real system,
$G\equiv(\omega-H)^{-1}$, in terms of the effective Green's
function, $\bar G$, and the total scattering matrix, $T$, as
\begin{equation}\label{GTG}
G=\bar{G}+\bar{G}T\bar{G}\  ,
\end{equation}
where, $T=V(1-\bar{G}V)^{-1}$. Using the multiple scattering
theory \cite{Gonis}, one can express $T$ as the multiple
scattering series
\begin{equation}\label{T}
T=\sum_it_i+\sum_i\sum_{j\neq i}t_i\bar{G}t_j+\sum_i\sum_{j\neq
i}\sum_{k\neq j}t_i\bar{G}t_j\bar{G}t_k+\cdots\ .
\end{equation}
Here, $t_i(\equiv t_{m,n})$ is the single-site $t$ matrix which
represents the multiple scattering of carriers due to the isolated
potential $v_i(\equiv v_{m,n})$ in the effective medium, and is
expressed as
\begin{equation}\label{ti}
t_i=v_i(1-\bar{G}v_i)^{-1}\  .
\end{equation}
The averaging of Eq.(\ref{GTG}) and the use of single-site CPA
condition $\langle t_i\rangle_\mathrm{av}=0$ for any site $i$ in
the wire, leads to $\langle T\rangle_\mathrm{av}=0$ and thus, we
obtain $\langle G\rangle_\mathrm{av}=\bar{G}$, as mentioned above.

The present system includes both substitutional disorder and spin
scattering. Therefore, the CPA equation for the coherent potential
is given by
\begin{equation}\label{14}
\langle
t_{m,n}\rangle_\mathrm{av}=(1-x)t_{m,n}^\mathrm{A}+x\langle
t_{m,n}^\mathrm{M}\rangle_\mathrm{spin}=0 \ ,
\end{equation}
where, $t^\mathrm{A}_{m,n}(t^\mathrm{M}_{m,n})$ represents the
complete scattering associated with the isolated potential
$v^\mathrm{A}_{m,n}(v^\mathrm{M}_{m,n})$ in the effective medium,
and $\langle\cdots\rangle_\mathrm{spin}$ denotes average over the
spin scattering at the M-site. In the classical spin treatment,
the potential for which a carrier is subjected at the M-site is
regarded as $\varepsilon_\mathrm{M}-IS$ or
$\varepsilon_\mathrm{M}+IS$, depending on whether the localized
spin on the M-site and the carrier spin are parallel or
antiparallel with each other. At the paramagnetic temperature at
which the orientation of localized spin is completely random, the
probability of each state is 1/2. Therefore, the associated
$t$-matrices for an arbitrary site of each atomic line, such as
$(0,n)$, are given by
\begin{equation}\label{15}
t_{0,n}^\mathrm{A}=\frac{\varepsilon_\mathrm{A}-\Sigma_n}
{1-(\varepsilon_\mathrm{A}-\Sigma_n)F_n} \ ,
\end{equation}
and
\begin{eqnarray}\label{16}
\langle
t_{0,n}^\mathrm{M}\rangle_\mathrm{spin}&=&\frac{1}{2}\left[\frac{\varepsilon_\mathrm{M}
-IS-\Sigma_n}{1-(\varepsilon_\mathrm{M}-IS-\Sigma_n)F_n}\right]\nonumber\\
&+&\frac{1}{2}
\left[\frac{\varepsilon_\mathrm{M}+IS-\Sigma_n}{1-(\varepsilon_\mathrm{M}+IS
-\Sigma_n)F_n}\right]\ ,
\end{eqnarray}
where
\begin{eqnarray}\label{17}
F_n(\omega)&=&\bar{G}_{n,n}(m,m;\omega)=\bar{G}_{n,n}(0,0;\omega)\nonumber \\
&=&\frac{a}{2\pi}\int_{-\pi/a}^{\pi/a}\bar{G}_{n,n}(k_x;\omega)\,dk_x
\mathrm{~~for~~} n=1,\cdots,N_y\nonumber\\
&&
\end{eqnarray}
In Eq. (\ref{17}), we should emphasize that the effective Green's
function, $\bar G$, depends on the one-dimensional wave vector
$k_x$ via $G^0$ (see Eq.(\ref{10})). Hence, to obtain any specific
matrix element $\bar G_{n_1,n_2}$, we must integrate over the
first Brillouin zone of the one-dimensional lattice \cite{Gonis}.

Equation (\ref{14}) leads to a cubic equation for the self-energy
of $n$th atomic line, $\Sigma_n(\omega)$, as
\begin{equation}\label{18}
\mathcal{A}\Sigma_n^3+\mathcal{B}\Sigma_n^2+\mathcal{C}\Sigma_n
+\mathcal{D}=0\mathrm{~~for~~} n=1,\cdots,N_y\  ,
\end{equation}
with
\begin{eqnarray}\label{19}
\mathcal{A}&=&-F_n^2 \\
\mathcal{B}&=&(\varepsilon_\mathrm{A}+2\varepsilon_\mathrm{M})F_n^2-2F_n \\
\mathcal{C}&=&[(IS)^2-2\varepsilon_\mathrm{A}\varepsilon_\mathrm{M}
-\varepsilon_\mathrm{M}^2]F_n^2\nonumber\\
&&+[x(\varepsilon_\mathrm{M}-\varepsilon_\mathrm{A})
+2(\varepsilon_\mathrm{M}+\varepsilon_\mathrm{A})]F_n-1 \\
\mathcal{D}&=&[\varepsilon_\mathrm{M}^2\varepsilon_\mathrm{A}
-(IS)^2\varepsilon_\mathrm{A}]F_n^2\nonumber\\
&&+[x(IS)^2+(x-2)\varepsilon_\mathrm{M}
\varepsilon_\mathrm{A}-x\varepsilon_\mathrm{M}^2]F_n\nonumber\\
&&+x(\varepsilon_\mathrm{M}-\varepsilon_\mathrm{A})+\varepsilon_\mathrm{A}\
.
\end{eqnarray}

It should be noted that, Eq. (\ref{13}) is a system of linear
equations which should be solved self-consistently, using Eqs.
(\ref{17}) and (\ref{18}), to obtain the coherent potentials
$\Sigma_n(\omega)$. Then, the LDOS per site in the $n$th atomic
line of the quantum wire, $g_n(\omega)$, is calculated by
\begin{equation}\label{20}
g_n(\omega)=-\frac{1}{\pi}\,\mathrm{Im}\,F_n(\omega)\  .
\end{equation}

We should remind the reader that, the LDOS is a function of the
energy and the space coordinate, which illustrates the spatial
distribution of the states at the particular location (here, $n$),
and it is well known that many important physical properties and
characteristics of a mesoscopic system are determined by the LDOS,
which is experimentally measurable.

When Eq. (\ref{18}) is solved for a real energy $\omega$, we
obtain three roots. We only choose the correct root corresponding
to $\omega+i\eta$, i.e., the imaginary part of $F_n$ must be
negative in order to give a positive LDOS. The existence and
uniqueness of such a solution depend on the initial guess for the
self-energies $\Sigma_n$. We believe that, the best guess for
$\Sigma_n$, is zero. One should note, however, that the real and
imaginary parts of the roots and the speed of convergence depend
on the parameters of the system. Except for a few values of the
energy, we have rapid convergence when the final value of
$\Sigma_n$ for a given energy is taken as the initial $\Sigma_n$
for the next energy. Furthermore, in the case of weak exchange
interaction, the convergence is faster than in the strong one.

It is important to note that, if we set $IS=0$, the present
formalism will be applicable for a NMS quantum wire of general
type $\mathrm{A}_{1-x}\mathrm{D}_x\mathrm{B}$ which has been doped
with donor or acceptor nonmagnetic impurities. Here, D indicates
the nonmagnetic impurity atom, and in the above equations, we
should set $\varepsilon_\mathrm{M}=\varepsilon_\mathrm{D}$. In
next section we present the numerical results of LDOS for both the
NMS and DMS quantum wires.

\section{Results and discussion}
In our numerical calculations, we measure the energies in units of
$t$ and we set $\varepsilon_\mathrm{A}=0$, since we can shift the
chemical potential without loss of physics. We present the
numerical results for both the NMS and DMS quantum wires with
$N_y=5$. In both cases, we have assumed the carrier density is
very low; hence, we have ignored the interaction between carriers.
In practice, we have done the numerical calculations for a case in
which only single carrier moves in the conduction (or valance)
band of the quantum wire.

In quantum wires, quantum effects influence the electronic
properties of the system. Due to the confinement of free carriers
in the transverse direction of the wire, their transverse energy
is quantized into a series of discrete values. In practice, when
the size or dimension of a material diminishes to the nanoregion,
the charge carriers begin to experience the effects of
confinement, meaning that their motion becomes limited by the
physical size of the region or domain in which they move.
\begin{figure}
\centering \resizebox{0.35\textwidth}{0.2 \textheight}
{\includegraphics{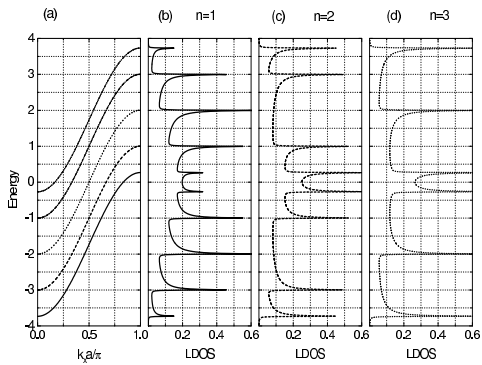}} \caption{The energy dispersion and
the LDOS curves of a clean semiconductor quantum wire. (a) The
dispersion curve as a function of normalized wave vector, and
(b)-(d) the LDOS as a function of energy, for the atomic lines
$n=1,5$, $n=2,4$ and $n=3$ respectively. Energy is measured in
units of $t$.}
\end{figure}

In Fig. 1, the energy dispersion and the LDOS curves of a clean
quantum wire are shown for comparison. In such system, there are
states with zero group velocity which are responsible for the
singularities of spectral density. Therefore, the van Hove
singularities appear as sharp features in the LDOS and due to the
presence of five atomic lines, five subbands are observed
\cite{Hugle}. Due to the symmetry of the system, the LDOS for
$n=1$ is the same as for $n=5$ (which are at the edges of wire),
and $n=2$ with $n=4$. Also, the number of sharp peaks depends on
the atomic line number.
\begin{figure}
\centerline{\includegraphics[width=1.1\linewidth]{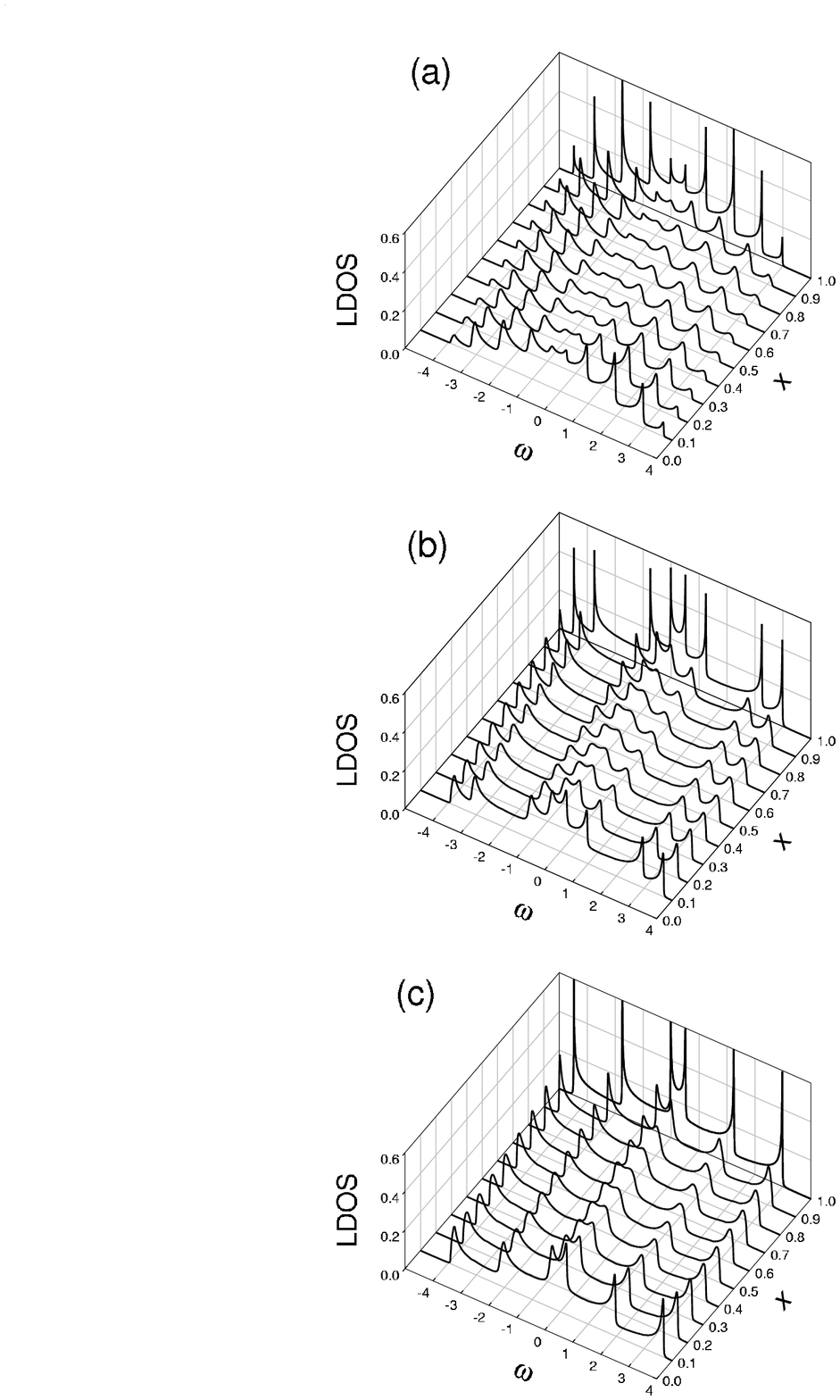}}
\caption{The LDOS of a NMS quantum wire with
$\varepsilon_\mathrm{D}=-0.75$ as functions of energy and impurity
concentration, for the atomic lines (a) $n=1,5$, (b) $n=2,4$ and
(c) $n=3$ respectively. Energy is measured in units of $t$.}
\end{figure}

To illustrate the impurity effects in NMS quantum wires, the LDOS
as a function of energy is plotted in Fig. 2 for
$\varepsilon_\mathrm{D}=-0.75$ and in Fig. 3 for
$\varepsilon_\mathrm{D}=+0.75$. An impurity with negative
(positive) site energy is analogous to an acceptor (a donor)
center at the host crystal. At zero concentration of impurity
($x=0$), the band for all atomic lines, is equivalent to the one
in the clean system (see Fig. 1(b)-1(d)). For the case of
$\varepsilon_\mathrm{D}<0$ ($\varepsilon_\mathrm{D}>0$), with
increasing $x$ the band shifts towards lower (higher) energies.
Also, with increasing $x$, the relative sharpness of peaks reduces
and at $x=0.5$ which represents the state of maximum
substitutional disorder, the LDOS is completely symmetric with
respect to the center of band and the sharpness disappears. With
further increasing $x$, the peaks again become sharp, and for
$x=1$, the LDOS with a shift equal to $\varepsilon_\mathrm{D}$, is
completely equivalent to the band of clean system. The reason of
such behavior is that, for $x>0.5$ the majority of atoms have
$\varepsilon_\mathrm{D}\neq 0$ ($\varepsilon_\mathrm{D}=-0.75$ or
$\varepsilon_\mathrm{D}=+0.75$ depending on the type of impurity).
Thus, one can imagine the situation in which the atoms with
$\varepsilon_\mathrm{D}=0$ act as impurity atoms in a host crystal
with $\varepsilon_\mathrm{D}\neq 0$. These features can be seen in
all atomic lines. Therefore, the nonmagnetic impurity can change
the behavior of LDOS and there is not band broadening in the NMS
quantum wires.
\begin{figure}
\centerline{\includegraphics[width=1.45\linewidth]{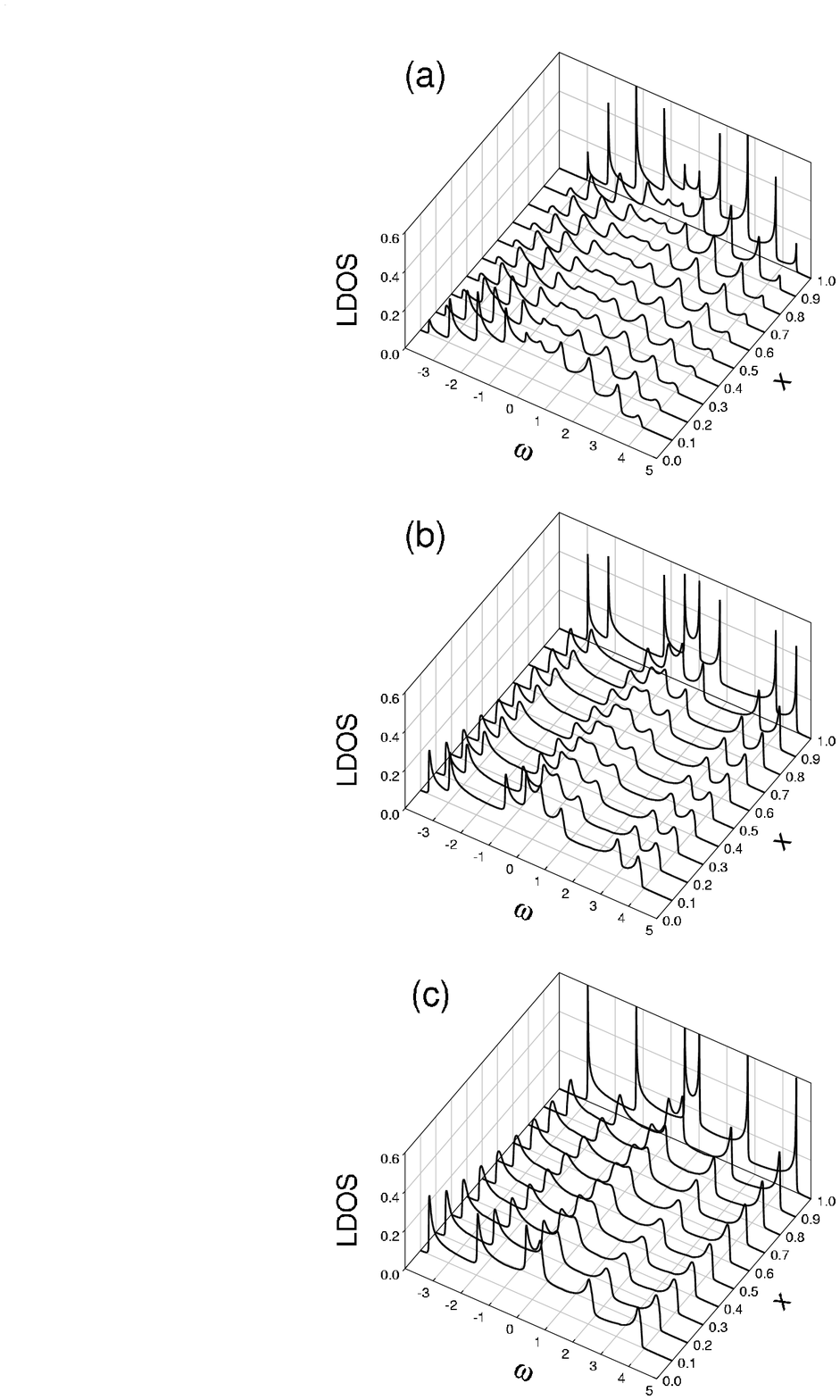}}
\caption{The same as Fig. 2 but for
$\varepsilon_\mathrm{D}=+0.75$.}
\end{figure}
We now investigate the influence of magnetic impurities on the
electronic properties of semiconductor quantum wires. Doping of
magnetic atoms such as Mn into GaAs or InAs quantum wires,
introduces not only magnetic moments but also free carriers.
Therefore, in such quantum wires, we should consider both the
effects of chemical potential and magnetic disorder on the LDOS at
the paramagnetic temperatures. Figures 4 and 5 show how the
carrier band changes with $x$ in the case of weak and strong
exchange interaction, respectively. We have shown the results for
$\varepsilon_\mathrm{M}=-0.75$, $IS=-0.4$ (weak exchange
interaction) and $IS=-1.2$ (strong exchange interaction) as
sampling cases of III-V-based DMS's. For this kind of impurity,
with increasing $x$ the sharpness of peaks decreases continuously
and at the concentration $x=1.0$ at which the LDOS is completely
symmetric with respect to the center of band, the van Hove
singularities completely disappear (particularly for $IS=-1.2$),
and the curves of all atomic lines become smooth. Such features
have not been observed in the bulk case \cite{Taka99}. It should
be noted that the results are same for different signs of $IS$,
because the system is in paramagnetic phase and the spin of
magnetic atoms is treated classically.
\begin{figure}
\centerline{\includegraphics[width=1.45\linewidth]{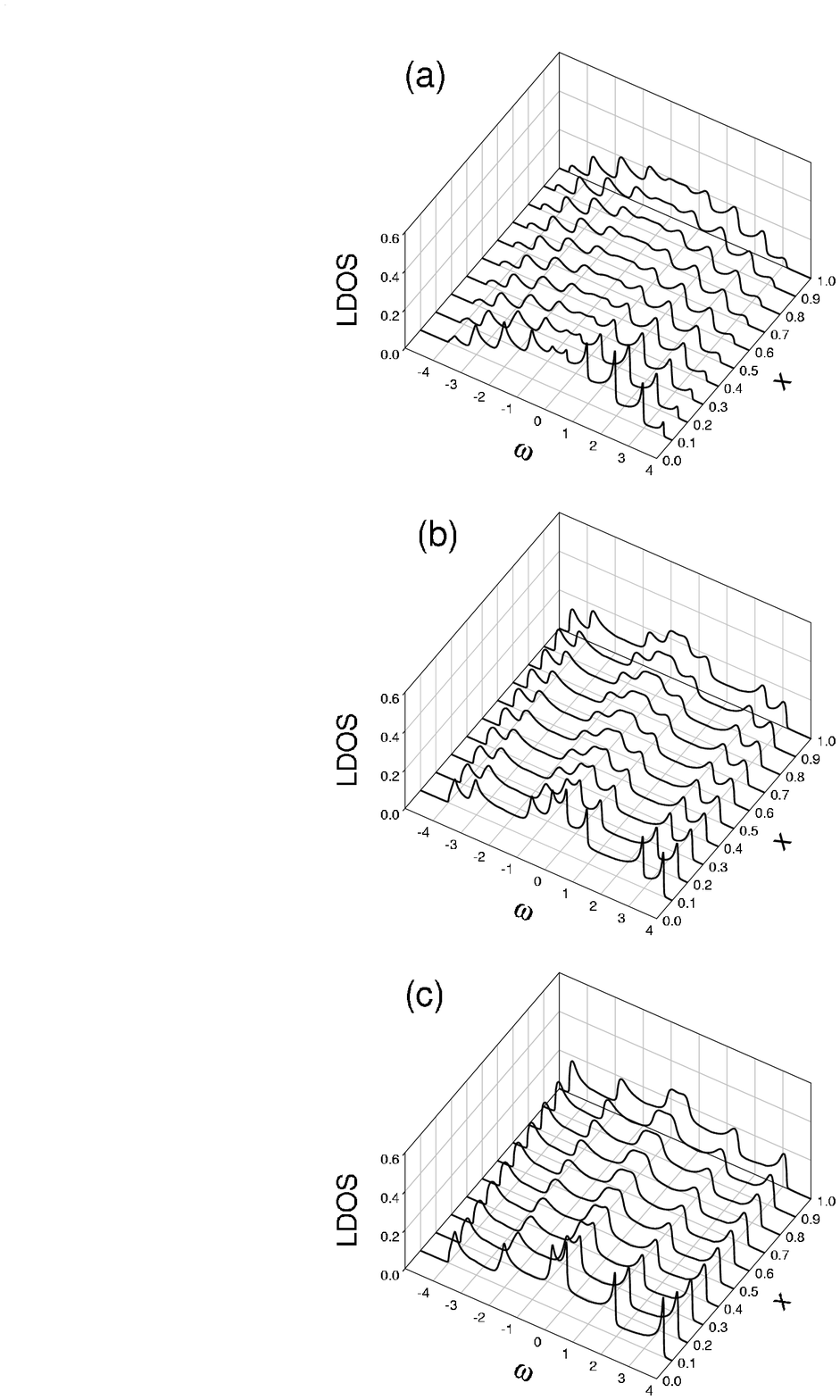}}
\caption{The LDOS of a DMS quantum wire with
$\varepsilon_\mathrm{M}=-0.75$ and $IS=-0.4$ as functions of
energy and impurity concentration, for the atomic lines (a)
$n=1,5$, (b) $n=2,4$ and (c) $n=3$ respectively. Energy is
measured in units of $t$.}
\end{figure}

On the other hand, the results show that, the LDOS of atomic lines
is different with each other. This feature indicates that the
electronic transport depends on the atomic line number. The
difference in the density of states of the atomic lines predicts a
nonuniform charge distribution in such quantum wires. Therefore
the atomic lines will have different contributions in carrier
transport and cause the quantum interferences, which can be
important in the process of charge transport through nano-scale
devices. Similar feature in the coherent electron conductance of a
quantum point contact in the presence of a scanning probe
microscope tip, has been reported both experimentally
\cite{Topinka} and theoretically \cite{He}.
\begin{figure}
\centerline{\includegraphics[width=1.45\linewidth]{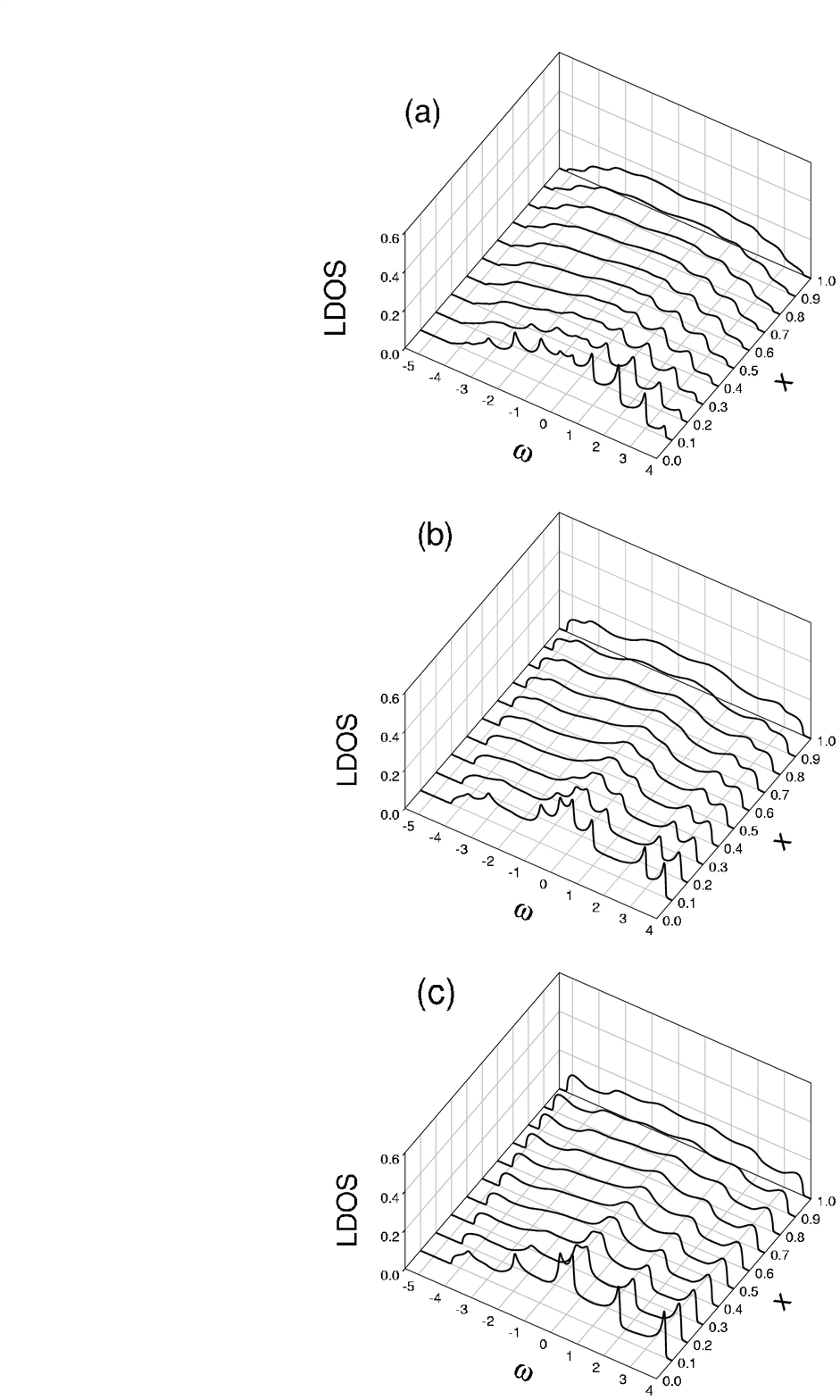}}
\caption{The same as Fig. 4 but for $IS=-1.2$.}
\end{figure}

Now, we discuss the band-edge energy shift and the band broadening
due to the random distribution of $\mathrm{M}$ ions and the
fluctuation of localized spins. The energy shift and the
broadening of the band can be calculated using the band-edge
energies, $\omega_b$, at which the density of states goes to zero
and the imaginary part of the self-energy vanishes. The
approximate solution for the lower (upper) band edge is given by
\cite{Taka99}
\begin{equation}\label{wb}
\omega_b=\Sigma_n^{l(u)}(\omega_b)\mp W_0\ ,
\end{equation}
where $\Sigma_n^{l(u)}(\omega_b)\equiv\Sigma_b^{l(u)}$ is the
energy shift of the lower (upper) band edge, and $W_0\simeq
3.74\,t$ is the half-bandwidth of the clean system. Using Eqs.
(\ref{17}) and (\ref{wb}), we numerically obtain
$F_n(\omega_b)t\simeq\mp 0.81$. Substituting these values into Eq.
(\ref{18}), we obtain real $\Sigma_b^{l(u)}$ or the energy shifts
of the band edges for the system. In Fig. 6, the band-edge energy
shift and the band broadening are depicted as a function of the
impurity concentration for various values of $IS$, with
$\varepsilon_\mathrm{M}=0$ for simplicity. In such a case, the
LDOS is symmetric around $\omega=0$, hence the magnitudes of the
band-edge energy shifts are equal and the band broadening is given
by $2|\Sigma_b^{l}|$ or $2|\Sigma_b^{u}|$. The results clearly
show that, in the presence of magnetic impurities, the bands are
broadened with the increase of $x$ due to the impurity spin
fluctuation. This band broadening depends on the strength of
magnetic disorder and is much larger when the interaction is
stronger (compare Fig. 4 and 5).
\begin{figure}
\centerline{\includegraphics[width=0.6\linewidth]{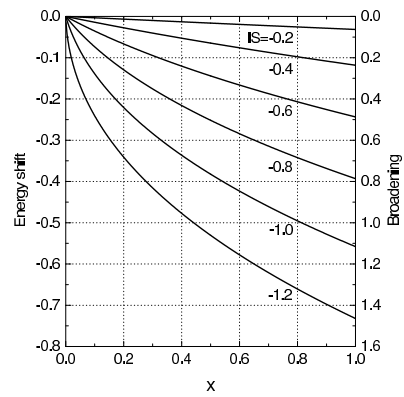}}
\caption{The energy shift of the lowest band edge and the band
broadening for various values of $IS$, with
$\varepsilon_\mathrm{M}=0$ as a function of impurity
concentration. Energy is measured in units of $t$.}
\end{figure}

We applied our theory for both the NMS and DMS quantum wires, by
employing parameters for simulating the semiconducting alloys. The
parameters of the tight-binding model used in this work can be
chosen from realistic orbitals of atomic species which are
suitable for an experimental realization of the modeled quantum
wires. Such parameters have been used to study bulk DMS's
\cite{Taka99,Taka02}. However, we should note here that, our
investigation is based on a single-band model and a single-site
approximation, neglecting many features such as multiband effects,
off-diagonal disorder, Coulomb interactions between free carriers,
and correlated defects. If such features affect the physical
properties of a real system, the above mentioned model should be
improved. Otherwise, one cannot expect to obtain accurate results
for that system, using the experimental parameters.

One could consider the present system as an infinite stack of
atomic slices along the $x$ direction. This means that we consider
the disorder in the $y$ direction. Thus, in such a case, we should
define a position-dependent self-energy for each slice. In other
words, the self-energy, in the cross section of the wire, is
different from one site to the other; however, the slices have
equal self-energy values. The single-site CPA condition should be
used for each slice to obtain the self-energies, and, in order to
derive the Green's function of such a system, one can use a method
similar to that for semi-infinite leads \cite{Nik1,Nik2,MacK}.

The calculations presented here can be extended for more than one
atomic strip and study the transport of free carriers parallel to
the strips (layers). In such a case, the transport properties of
the system depend on the strip number, which can be attractive for
planar devices with submicron dimensions.

\section{Conclusions}
We have studied the effects of magnetic and nonmagnetic impurities
on the electronic states of semiconductor quantum wires. Using the
CPA for random distribution of impurity atoms, we investigated the
influence of impurity concentration and the strength of exchange
interaction on the LDOS. In NMS quantum wires, the acceptor
(donor) impurities only shift the bands towards lower (higher)
energies, and the van Hove singularities in the LDOS depend on the
value of impurity concentration. For DMS quantum wires, the
magnetic impurities broaden the bands and reduce continuously the
peaks sharpness, even in the case of weak exchange coupling. The
results presented here predict that the DMS quantum wires can be
used in laser operation and the physics of spintronic nanodevices.

\section*{Acknowledgments}
The author would like to thank Professor R. Moradian for helpful
discussions. This work was supported by the Payame Noor
University.

\end{document}